# Local ultra-densification of single-walled carbon nanotube films: modeling and experiment

*Artem K. Grebenko\*, Grigorii Drozdov, Yuriy G. Gladush, Igor Ostanin, Sergey S. Zhukov, Aleksandr V. Melentyev, Eldar M. Khabushev, Alexey P. Tsapenko, Dmitry V. Krasnikov, Boris Afinogenov, Alexei G. Temiryazev, Viacheslav V. Dremov, Traian Dumitricã\*\*, Mengjun Li, Hussein Hijazi, Vitaly Podzorov, Leonard C. Feldman, Albert G. Nasibulin\*\*\**

Dr. A.K. Grebenko, Prof. Y.G. Gladush, Prof. I. Ostantin, E.M. Khabushev, Dr. A.P. Tsapenko, Dr. D.V. Krasnikov, A.G. Nasibulin
Skolkovo Institute of Science and Technology, Nobel str. 3, 121205, Moscow, Russia.

Dr. A.K. Grebenko, Dr. S.S. Zhukov, A.V. Melentyev, Dr. V.V. Dremov
Center for Photonics and 2D Materials, Moscow Institute of Physics and Technology, Institute Lane 9, Dolgoprudny, Russia.

Dr. G. Drozdov, Prof. T. Dumitricã
Scientific Computation Program, University of Minnesota, 55455, Minneapolis, USA

Prof. I. Ostanin
University of Twente, PO Box 217, 7500 AE, Enschede, the Netherlands

Dr. A.P. Tsapenko
Aalto University, Department of Applied Physics, Puumiehenkuja 2, FI-00076 Espoo, Finland

Dr. B. Afinogenov
Faculty of Physics, Lomonosov Moscow State University, Leninskie gory 1, Moscow, Russia

Dr. A.G. Temiryazev
Kotel'nikov Institute of Radioengineering and Electronics of RAS, Fryazino Branch, Vvedensky Sq. 1, Fryazino, Russia

Dr. V.V. Dremov
Russian Quantum Center, Bolshoy Boulevard 30 s1, Moscow, Russia

Prof. T. Dumitricã

Department of Mechanical Engineering, University of Minnesota, 55455, Minneapolis, USA

Dr. M. Li, Dr. H. Hijazi, Prof. V. Podzorov, Prof. L.C. Feldman

Department of Physics, Rutgers University, Piscataway, New Jersey 08854, USA.

Dr. M. Li

Intel Corporation, Albuquerque, New Mexico.

Prof. A.G. Nasibulin

Aalto University, P.O. Box 16100, FI-00076 Aalto, Finland

E-mail: *artem.grebenko@skoltech.ru, **dtraian@umn.edu, ***a.nasibulin@skoltech.ru

Fabrication of nanostructured metasurfaces poses a significant technological and fundamental challenge. Despite developing novel systems that support reversible elongation and distortion, their nanoscale patterning and control of optical properties remain an open problem. Herein we report the atomic force microscope lithography (AFML) application for nanoscale patterning of single-walled carbon nanotube films and the associated reflection coefficient tuning. We present models of bundling reorganization, formed-pattern stability, and energy distribution describing mechanical behavior with mesoscopic distinct element method (MDEM). All observed and calculated phenomena support each other and present a platform for developing AFML patterned optical devices using meshy nanostructured matter.

**1. Introduction**

Atomic force microscopy (AFM) evolved from a visualization device to an instrument that addresses state-of-the-art fundamental problems,[1] measures various material properties,[2] and facilitates the creation of nanostructures.[3] AFM-based lithography (AFML) is a promising tool for high-resolution (~ 5 nm) device fabrication[4] through indentation,[5] electrochemical anodic oxidation,[6] and "ink" deposition[7] using the AFM tip to modify the target surface. For large-throughput applications, the cantilever can be extended to an array of indenters.[8,9] Recently, AFML has been applied to 2D materials facilitating the fabrication of single-photon emitters with exceptional precision.[10,11] In contrast to optical (OL) or e-beam (EBL) lithography, AFML usually does not require an intermediate resist film for patterning. The absence of such additional layers is critical when the substrate is mechanically unstable (i.e., resist application induces material modification) or when the target surface is small or curved. Compared to OL and EBL, AFML facilitates patterning on top of polymer films that do not withstand harsh organic solvents and baking procedures.

One of the most striking applications of EBL and OL is the fabrication of metasurfaces, a subwavelength layer designed to form a wavefront of interest.[12,13] Many prototypes were fabricated with these approaches, including flat lenses and axicons,[14,15] vortex generators,[16] polarization converters and splitters,[17,18] perfect absorbers and modulators,[19] as well as holograms.[20,21] The extension of such systems to small and curved surfaces can be achieved, e.g., by the assembly of nanomaterials: nanospheres[22–24] and single-walled carbon nanotube films (SWCNT) films,[25,26] which could be further modified with the help of AFML. However, in this instance, regular AFML is also limited by mechanical instability of nanostructured ensembles that are challenged during lateral movements of the tips when brought into direct contact. Here we implement a method of AFM indentation free from these restrictions. Through tip pressing, we locally change the density of the nanotubes in the film, thus controlling its optical properties.

We have chosen SWCNT films as a model system thanks to multiple state-of-the-art practical applications for conductive, flexible, and stretchable optoelectronic devices.[25,27–29] Being a sparse network of randomly oriented nanotubes, SWCNT film is mechanically unstable if, for example, a liquid is applied to the surface[30] (which alters their optical properties[31]) or when an AFM tip is laterally moving in direct contact with the film. Previously the control of optoelectronic properties of SWCNT films was performed by means of OL and EBL followed by oxygen plasma etching[32] and lift-off routines.[33] Lithography-free modification of optoelectronic properties is also reported with the densification driven by interaction with liquid[30] or by alignment of nanotubes.[34] These methods allow the creation of electromagnetic shielding[35] and reflective surfaces[36] in macro-size samples. Another strategy is to employ pre-patterned growth of carbon nanotubes forests.[37–39] However, there are no reports of local density control in SWCNT films or other meshy structures with a sub-micron resolution and precision. Therefore, a method that would deliver precise control of optoelectronic properties of meshy structures at nanoscale resolution, combined with the latest advances in flexible and stretchable metasurfaces, is highly desired.[40–42]

Here, we propose an experimental method of nanoscale patterning of SWCNT films to control their optical properties with the help of a novel AFML technique. We present a careful spectroscopic analysis and several proof-of-the-concept devices. Mesoscopic modeling of individual indentation events is used to reveal the microstructural changes occurring inside the SWCNT film under the tip pressure. Altogether, we present the first nm-scale dry patterning technique for meshy nanostructured ensembles and a platform to analyze and predict their mechanical properties.

## 2. Results

SWCNTs were synthesized using an aerosol (floating-catalyst) chemical vapor deposition technique, discussed in details elsewhere.[43,44] Briefly, SWCNT aerosols were deposited on nitrocellulose filters as randomly oriented nanotubes and subsequently dry-transferred to a target (quartz, $CaF_2$, and $Si/SiO_2$) substrate.[45] Pristine, i.e., as-produced SWCNT films were highly sparse with negligibly small reflectivity, while the remains of catalytic particles have a negligible effect on optical properties of the film.[46] To analyze systems on the scale of individual nanotubes we employ computer-generated simulation since such experimental analysis is a challenging task considering meshy samples. To address this issue, we used the Mesoscopic Distinct Element Method (MDEM).[47] The modeling of SWCNT materials in given time and length scales requires advanced coarse-graining techniques. The MDEM model for SWCNTs interaction,[48] informed by atomistic modeling[49] and highly parallelized for distributed memory computational clusters,[47] provides optimal accuracy for targeted applications.[30]

In our samples, nanotubes are oriented randomly but mostly in-plane, so we perform indentation in the perpendicular direction to nanotubes' long axes. The dry densification process consists of repeated push-pull-shift sequential steps performed with an Atomic Force Microscope (AFM) using two different regimes of microscope operation. The first one (implemented as PeakForce™ tapping and HybriD™ mode) is based on the tip touching the sample surface in each scan point, while moving in lateral directions in a retracted state. Each time the complete force-distance curve is measured using the load force value as a feedback parameter. The second regime, the hard-tapping, is based on the tip oscillating at its resonant frequency and the microscope feedback controlling the set-point of the oscillation magnitude. In contrast to the liquid-driven densification,[30] the dry approach enables patterning density-templates at micro- and nanoscale with a high localization precision (minimal pixel position accuracy) and resolution (minimal pixel lateral size) which falls in the range of 5-20 nm for regular AFML.[50] However, in this case, the term resolution is specific to a particular sample and tip as they both limit the force-driving printing resolution. As for the sample, the resulting crater shape is influenced by the sample thickness during the indentation procedure. At the same time, the indenter shape, which is not an ultrathin rod for regular AFM tips, also inevitably affects the created deepening. We used the usual pyramidic tips and SWCNT films with thicknesses varying from 100 to 1500 nm, therefore achieving the resolution in the range from 50 to 250 nm with localization precision of approximately 20 nm.

The process is illustrated with the help of MDEM simulations of a computer-generated SWCNT film sample (**Figure 1A**). In the first step, the sharp AFM tip is pushed into the film in a velocity-controlled manner until the desired force is reached (**Figure 1B**). The force value is set either directly or through a set-point value of the tapping amplitude. Next, the tip is retracted (**Figure 1C**). Once no longer in contact with the film surface, the tip undergoes a 10 nm lateral shift. This 3-step process is then repeated (**Figure 1D, E**), ultimately leading to the film densification. For example, after about a million nano indentations, we have densified in such a manner prominent (10x10 µm$^2$) square areas (see Materials and Methods section). Zoomed surface regions captured by Helium Ion Microscopy (HeIM) of the films densified under different force values are shown in **Figure 1F-K**. The degree of densification is quantified in **Figure 1L** by showing the relative thickness $h/h_0$ ($h$ - relaxed film thickness, $h_0$ - pristine film thickness) of the densified regions as measured in a large set of experiments carried out on pristine films with the same AFM tip but different applied force values. These figures demonstrate that larger applied forces are associated with smaller pore sizes of the film microstructure and smaller relative thickness values. The MDEM simulations displayed in **Figure 1D** and **E** reveal that the subsequent push-pull process does not affect the depth of the well created in the previous process. Thus, the height of the film densified by repeated applications of this process could be accurately approximated by h. Indeed, in comparison with experimental data presented in **Figure 1L**, the $h/h_0$ (ratio between relaxed film and pristine thickness) computed for a series of applied stress values follow the experimental data closely while the $h_m/h_0$ (ratio between maximal displacement and pristine thickness) rather overestimates the relative height reduction of the film.

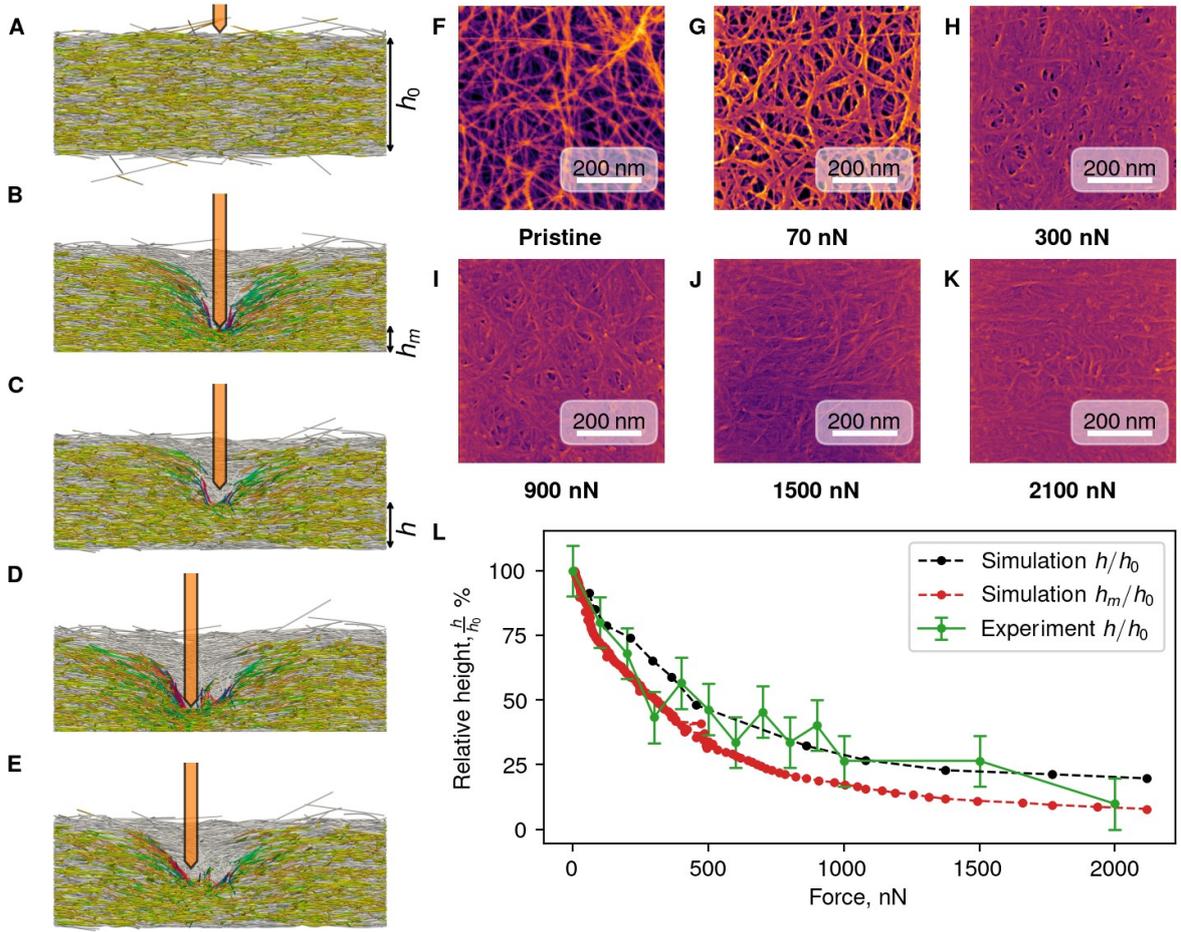

**Figure 1. Densification of SWCNT film with a nanoindenter. (A-E)** MDEM simulations showing densification mechanism. The middle cross-section of (**A**) computer-generated pristine 0.3x0.3 µm². SWCNT film sample with with initial thickness of $h_0$ = 127 nm at different stages of indentation: (**B**) densified by 20x20 nm² indenter down to the thickness of $h_m$=25 nm, (**C**) recovered to the thickness of $h$, (**D**) densified with the indenter shifted by 10 nm to the left, (**E**) recovered from the second indentation. Colors show the vertical orientation of SWCNT distinct elements. (**F-K**) HeIM images of the film regions densified by a 28 nm sharp AFM tip. (**L**) Comparison of relative height $h/h_0$ versus densification force for experimental results (AFM) and simulations for as-compressed film and after recovery.

The MDEM simulations summarized in **Figure 2** describe the microstructural changes in the SWCNT film during the push-pull step. **Figure 2A** displays the pristine SWCNT film (prior to nanoindentation) exhibiting a complex microstructure that reflects the balance between the SWCNT elasticity and the inter-tube van der Waals (vdW) forces. The attractive component of the vdW drives SWCNT aggregation into bundles, while the repulsive component causes entanglement, which hinders film densification.[51] As such, the resulting stabilized film comprises SWCNTs organized into dendritic bundles.[30] Entanglement prevents the thinner bundles that are branching out, visible in the dendrite agglomerate (maroon SWCNTs) of **Figure 2A** callout, from aggregating into a single thicker bundle. As noted from **Figure 1A-C**, the individual indentation process is creating a localized plastic deformation in the form of a deep well. We have monitored the stress applied through the nanoindenter during the push, **Figure 2B**, as well as the changes in vdW and strain energy stored by the SWCNTs of the whole film sample, during the push (**Figure 2C**) and pull (**Figure 2D**). The stress-strain curve shown in **Figure 2B** reveals that an initially soft compression regime, for strains less than ~25 % and stress values below 100 MPa, is followed by a hardening regime, where stress rises to several GPa. In the soft regime, the changes in film total energy are negligble, as the lowering of the vdW energy is compensated by a slow monotonic increase of the strain energy component, which represents the SWCNT bending (**Figure 2C**). The vdW energy gain occurs because SWCNT segments located underneath the indenter are being brought into closer contact, minimizing the voids in the initial microstructure (**Figure 2E**). It is worth noting that pulling out the indenter at this stage does not lead to permanent structural changes. After ~25 % compression, the strain energy increases significantly while the vdW energy continues to drop (**Figure 2C**). The strain energy increase is conditioned chiefly by bending the SWCNTs located below the indenter. At the microstructural level (**Figure 2E** and **2F**) we observed restructuring in the region under the indenter through the aggregation of new aligned SWCNTs (red-colored SWCNTs) and coarsening of the existing bundles (maroon SWCNTs). Such changes demonstrate that the entanglement established in the original film has been altered. At maximum penetration, where the distance from the indenter to the film lower surface hm is only 25 nm ($h_m/h_0$ ~ 10 %), we found that the total energy of the film increased by 16 eV/SWCNT.

On the pull-out step, modeled here as the evolution of the SWCNTs without the externally applied pressure, the film displays a pronounced plastic response. During the first ~ 5 ns, the evolution of the film microstructure is driven by the strain energy release, shown in **Figure 2D**, which leads to a lifting of the well bottom. However, this tendency is countered by the gained vdW adhesion from the bundle portions formed during the push step. After ~5 ns, the film dynamics slows down; its evolution is now driven by a tendency to increase the vdW adhesion through more bundling. At the end of the pull-off process, the well depth reaches an *h* value of 41 nm, which is 68% lower than the original $h_0$. Concerning the film at maximum penetration, the pull-out stage lowers the total energy by ~33 eV/SWCNT with respect to the maximum penetration configuration. In **Figure 2H**, it is interesting to note that when compared to the original pristine film, the locally-densified film has more bending strain energy, which is strongly localized at the well. However, the vdW energy (**Figure 2I**) is also significantly lowered around the well (**Figure 2G**). In spite of the augmented strain energy penalty, the substantial gain in vdW energy renders the locally deformed film by 17 eV/SWCNT lower in energy than the pristine one. A similar conclusion can be drawn for the film subjected to equispaced indentations (**Figure 2J**). For example, we found that when the film is subjected to 121 indentations, it densifies into a 0.25 g/cm$^3$ entangled structure with 3.4 eV/SWCNT lower energy than the pristine one (**Figure S1**).

As seen above, the MDEM simulations predict that the films densified in this manner exhibit superior stability as there is no energetic advantage to recover in the initial low-density pristine state. Moreover, when subjected to lateral strains, the simulated strain-stress curves of **Figure 2K** indicate that the densification process is not deteriorating the stretch mechanics. The pristine film exhibits significant plasticity above ~0.5 % strain, where the steep rise of the stress with a Young modulus of 6.9 GPa is followed by pronounced flattening, which is associated with the partial loss of the original entanglement and SWCNT sliding.[51] The densified film displays a similar Young modulus of 7.9 GPa, indicating that in this configuration SWCNTs still present significant entanglement. Note that as they cross multiple well regions, the SWCNTs of the densified film (**Figure 2J**) display waviness, an effect that promotes entanglement.[49] Figure 2K also indicates that the entanglement of the densified film is sufficiently robust to inhibit plasticity. As a result, the stress-strain curve maintains a monotonically increasing dependence steeper than the one calculated for the originally pristine film.

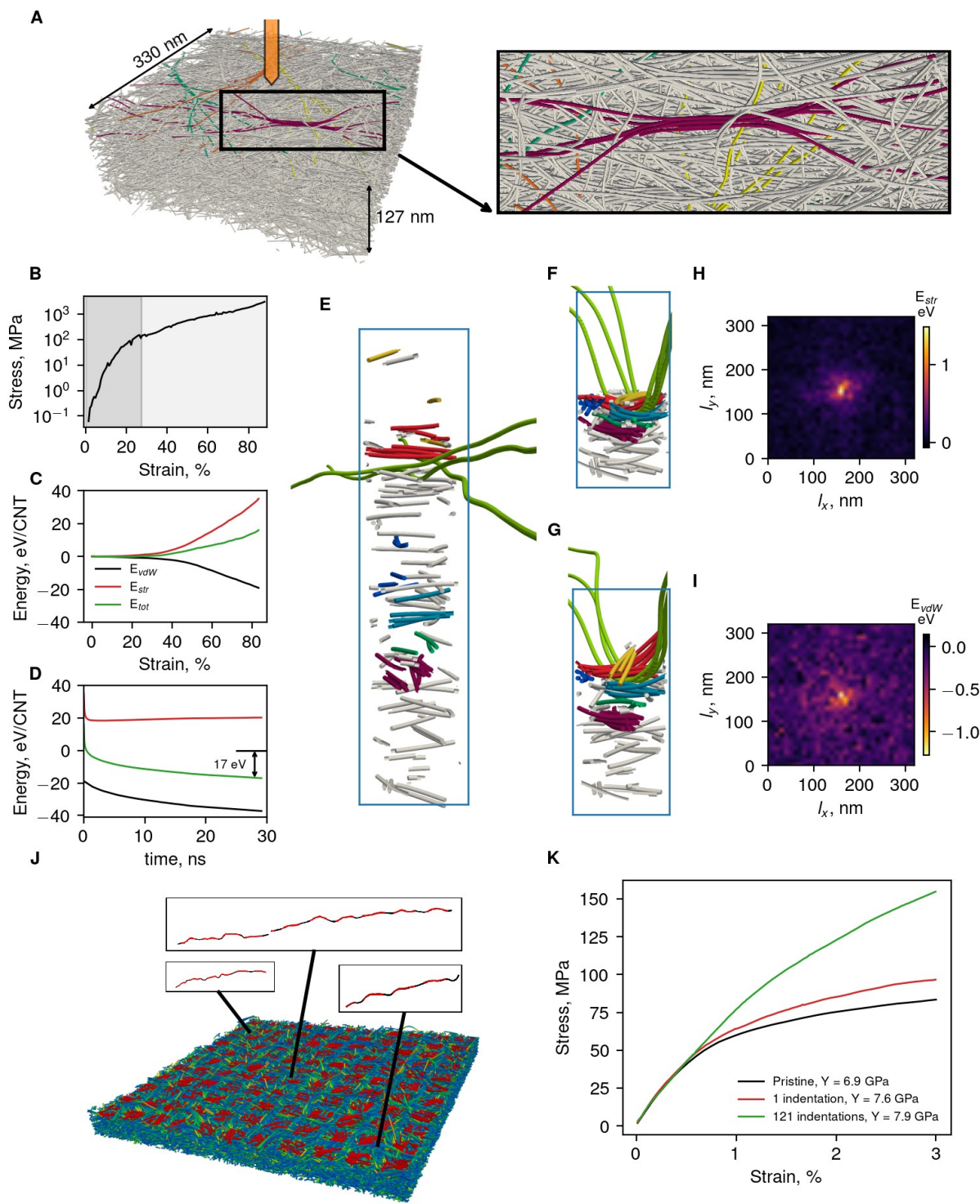

**Figure 2. Computational investigation of SWCNT film's structural properties during single indentation.** (**A**) Overall view of the simulated film. Colors are used to show branching structure of the bundles. The callout shows a selected dendritic bundle. (**B**) Stress-strain curve on a logarithmic scale during the indenter push step. (**C-D**) Change of vdW, strain, and total energies (**C**) during indenter push with respect to the pristine film state and (**D**) during indenter pull out with respect to the maximum compressed film state. The legend is the same for both panels **C** and **D**. (**E-G**) Microstructure evolution of the film located directly under the indenter (**E**) at the initial stage, (**F**) maximum penetration of the indenter, and (**G**) full film recovery (complete pull out). Colors are used to indicate SWCNTs that aggregate to form new large bundle portions in (**G**). (**H-I**) Maps of (**H**) strain and (**I**) vdW energy changes for the locally-densified film with respect to the pristine one. (**J**) The film under 11x11 equispaced indentations at maximum penetration. The callouts show the waviness feature of selected SWCNTs. (**K**) Stress-strain curves for lateral stretching of the pristine and densified film with single and multiple indentations.

We experimentally consider large (50-3000 µm) samples studied using broadband spectroscopy. These investigations indicate that the dry densification process does not alter some intrinsic properties of the SWCNTs, which fully coincides with computational results. The transmission coefficient does not significantly change in the whole investigated range, as illustrated in **Figure 3A** and **Figure S2**. UV-vis-NIR spectroscopy indicated typical absorption lines, i.e., transition between van Hove singularities, attributed to the distribution of various chiralities and diameters. As well, electrical conductivity determined from THz spectroscopy measurements (**Figure S2**) revealed minor changes in the free-electron dynamics after the densification. Raman spectroscopy (**Figure 3B**) demonstrates the preservation of SWCNT characteristic bands during the densification (for individual spectra see **Figure S2**). The intensity of the D-band compared to the most intensive G-band insignificantly increases as the densification force values increase. Therefore, the densification process does not significantly increase the defects of the individual nanotubes within the film. At the same time, 2D-band intensity increases more than two times than the G-band, reaching $I_G/I_{2D} = 15$, and may be indicating the slightly increasing alignment of the nanotubes,[52] which can be a consequence of the increased bundling. Bundling enhancement is also consistent with the 1584 cm$^{-1}$ shoulder intensity increase compared to the G-band[53–55] (see **Figure S2**). Applied densification forces result in the broadening and intensity redistribution of the radial breathing modes (RBMs)[55](see **Figure S2**), which might be indicative of the fact that nanotubes with different diameters react differently to the densification procedure.

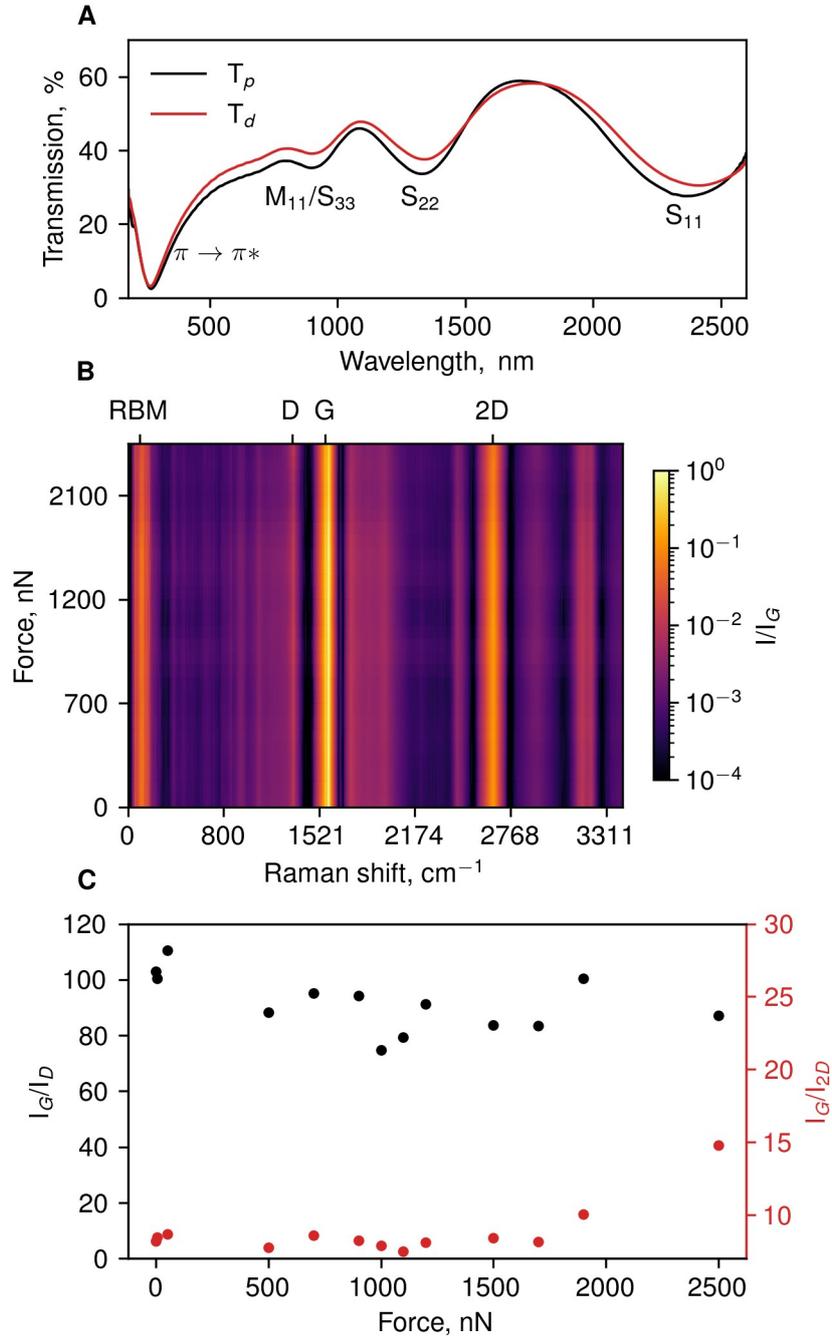

**Figure 3. Spectroscopy results confirming the preservation of the SWCNT structural properties. (A)** UV-vis spectroscopy of pristine film (40%) and densified by 2500 nN force. **(B)** 2D map indicating overall perseverance of typical Raman bands. The color indicates the intensity normalized per G-band. **(C)** Dependencies of the $I_G/I_D$ and $I_G/I_{2D}$ ratios on the densification force.

Due to the change of bundling and compression of the SWCNT film on various concrete substrates (Si/SiO$_2$, Quartz, CaF$_2$), densified regions tend to have significantly altered reflection coefficients, as presented in **Figure 4**. We will denote the reflectance of densified nanotube mats normalized per the reflectance of pristine film of a given initial thickness as reflection enhancement (RE). For the fabrication and measurements details, please refer to the Materials and Methods section. RE largely depends on the densification force and wavelength, as clearly seen from the maps illustrated in **Figure 4A** (indicated by color). Note that the change in the absolute reflection coefficient when normalized per quartz substrate or golden mirror is not prominent (see **Figure S3**).

The relative reflection change can reach two orders of magnitude. A set of maxima observed on this curve is associated with features in the SWCNT density of states, so-called Van Hove singularities. We analyzed the UV-THz range and found out that the most prominent phenomenon of RE occurs in the UV-vis-NIR range, and the shorter wavelengths lead to higher reflection enhancement on a par (**Figure 4B**). The maximum of reflection enhancement (MRE) for the vis-NIR range depends on the film's thickness (often presented in the form of the $T_{600}$[56]), as shown in **Figure 4C**. Assuming a slight change of the transmission coefficient, we associate reflection enhancement with lowering of the scattering (see **Figure 4D**) in the SWCNT films caused by bundling and overall inter-bundle distance decrease. For the pristine film, which is the sparse network without a well-defined film-air boundary, the back-reflection is dominated by the scattering. In contrast, the densified film possesses an abrupt flat air-film border, giving rise to specular reflection build-up. For the middle- and far-infrared ranges, the effect is much less prominent and can even be opposite – densified films reflect less when compared to the pristine films values (**Figure S3**). It should be also stressed, that when large reflective surfaces are considered, RE strongly depends on the density of the indentations (see **Figure S3**).

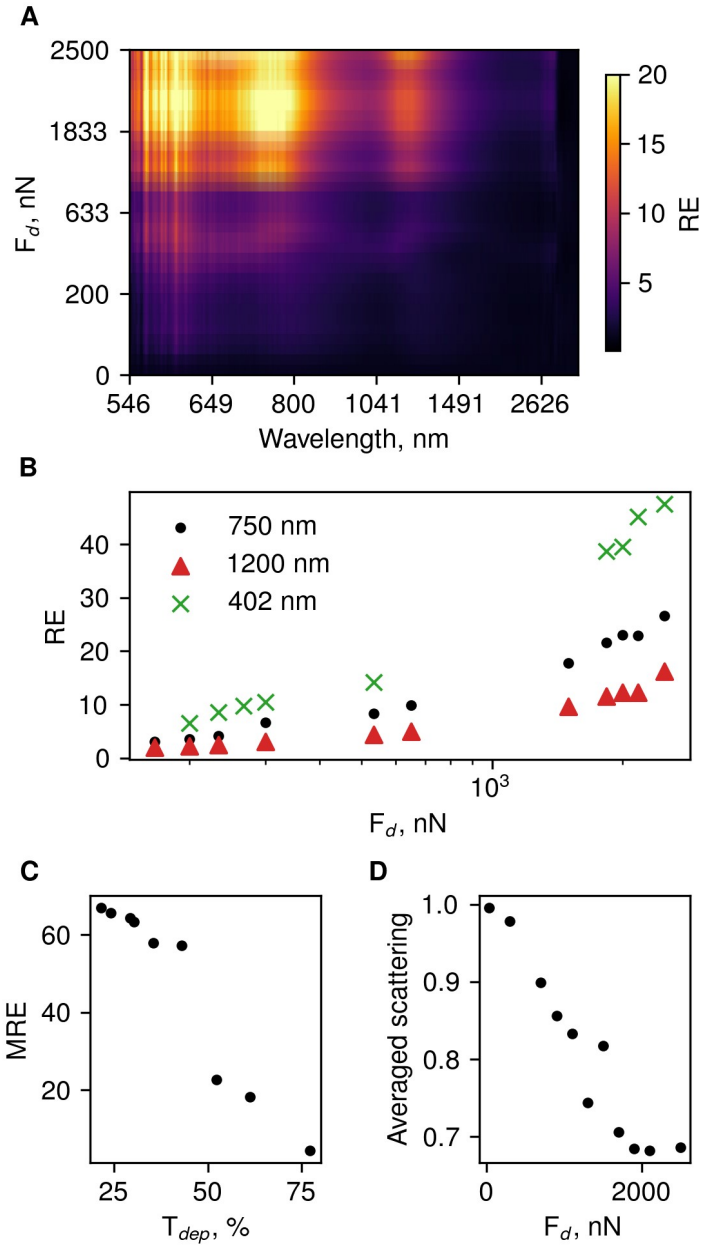

**Figure 4. Optical properties controlled by local densification. (A)** Reflection enhancement map illustrating dependence of RE on the densification force and wavelength. **(B)** Dependence of the reflection enhancement value normalized to the reflection of pristine film for UV, vis and NIR regions for a fixed thickness of the SWCNT film at ~ 250 nm ($T_{600}$ = 40-50 %). **(C)** Maximal Reflection Enhancement (MRE) for samples densified at 2500 nN of the different initial thicknesses ($T_{600}$). **(D)** Averaged scattering intensity dependence on the densification force, indicating one source of the reflection enhancement.

With this understanding of the optical properties, we can now focus on nanoscale patterning. In order to demonstrate the applicability of our approach, we have created several illustrative samples, such as computer-generated holograms (CGH) and diffraction gratings (summarized in **Figure 5**). CGH was calculated from the "Sk" letters image (**Figure 5A**) and later transferred by the AFML (Hybrid mode) to the 1.5 µm thick SWCNT film placed on top of a quartz substrate (**Figure 5B, C**). A thick sample was chosen to maximize the reflection contrast between pristine and densified regions. The size of the hologram was 100x100 µm2 with each image pixel having approximately 780 nm size. For each pixel we had 64 individual indentations operating at the resolution limit of the method. The correctness of the lithographic transfer was confirmed first by a Fourier transform of the SEM image (**Figure 5D**). Further, the holographic image was successfully reconstructed by a 650 nm laser (**Figure 5E**) on a white paper screen. This proof-of-concept demonstration allows us to be optimistic about the future applications of the proposed techniques in more complex optically active systems. More importantly, this hologram remained stable on the timescale of a few months without any noticeable degradation or relaxation of the densified nanotubes back to their pristine state.

By now, neither PeakForce™ nor HybriD™ modes are common, while the tapping regime is available in almost every microscope. We managed to successfully implement the above densification technique in this regime (see Materials and Methods for more details). We have defined a set of diffraction gratings with different periods ranging from 800 to 1800 nm with 100 nm increment, while keeping the duty fixed at 50 % (**Figure 5F-H**). These gratings clearly exhibit diffraction of the laser radiation with several observable maxima. We have performed a broadband spectroscopy of the formed structures and did not find any prominent difference in the whole range but from 1 to 3 µm (**Figure 5H**). The observed peak in the reflection spectra is not associated with the densification procedure, substrate, or the properties of the pristine films. However, neither the shift of the peak position nor the decrease of its intensity in the measured range allowed us to interpret the peak as some sort of collective excitations, such as, e.g., plasmon.

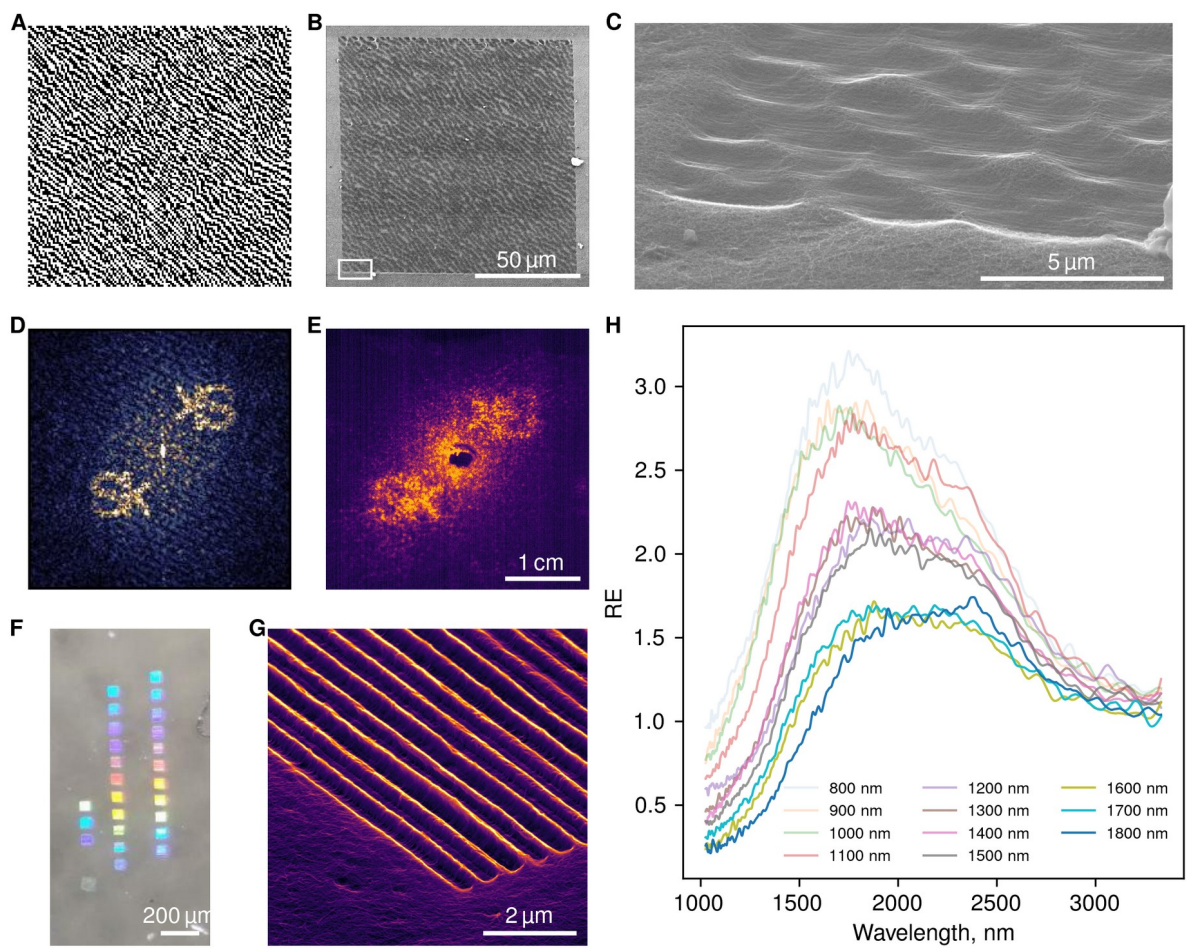

**Figure 5. Optically active media defined by means of AFML. A-E** Static CGH patterning in the SWCNT film on a quartz substrate. **(A)** Initial pattern generated by computer. **(B)** An SEM image of the pattern 120x120 µm² transferred by means of AFML (Hybrid™ mode) and **(C)** close look on the left bottom corner. **(D)** Fourier transform of image **(B). (E)** Optical photograph of the hologram recovered by 650 nm laser. **(F)** Optical photograph of a set of diffraction gratings with different periods patterned in SWCNT film on quartz substrate by hard tapping AFML. An SEM **(G)** image of one of the diffraction gratings. **(H)** NIR spectrum of the reflection enhancement of diffraction gratings indicating spectral features resulting from the pattern. This feature may be assigned to some collective excitations occurring in these gratings.

## 3. Conclusion

To sum up, we have developed a novel grayscale, dry lithographic control of optical properties of SWCNT film. We observed up to 50 times increase of the reflection coefficient relative to the pristine nanotubes at roughly 150 nm resolution when 2500 nN load force is applied. Such a method can be further utilized for the fabrication of a variety of stretchable and flexible optoelectronic devices in the meshy nanostructured matter. These results were supported with MDEM simulations that reveal the local meso-scale restructuring details responsible for the film densification. Simulations also indicated the mechanical stability of the formed pattern and its elongation properties. Overall this study presents a platform to predict and fabricate state-of-the-art devices employing peculiar optical properties of nanomaterials.

## 4. Methods

*Sample synthesis*: SWCNT films were synthesized using a floating catalyst (aerosol) chemical vapor deposition method in a vertical hot-wall tubular reactor[57,58] heated to 850 ℃. Ferrocene was used as a catalyst precursor, which vapor was delivered from a saturator (cartridge containing ferrocene and granular silica mixture) to the reactor hot zone by a gas flow. Carbon monoxide was utilized as a carbon source and a gas carrier simultaneously, while a small amount of CO2 (~1 vol%) was added into the reactor for nanotube growth promotion.[44] SWCNTs were grown in the aerosol phase and collected downstream of the reactor using a nitrocellulose filter and afterward transferred to the target substrate through the dry transfer technique:[45] a filter with SWCNT film was applied to the surface of the substrate, dry-pressed, and released.

*Atomic Force Microscopy Densification and Lithography*: For AFM-driven densification, we used Bruker Multimode V8 microscope operating in PeakForce tapping regime, which allowed a precise definition of the pressure value and absence of lateral deformation during densification as between pixels the tip moved in a retracted state. We have used RTESPA-300 and RTESPA-525 silicon probes for most densification experiments. They were calibrated using Thermal Tune and collecting force-distance curves on a reference sapphire sample. According to the procedure described elsewhere,[59] their sharpness was analyzed utilizing a dimpled aluminum sample fabricated at Moscow State University. These probes were ultrasharp, but they could not withstand the whole range of forces used for densification. In order to overcome this issue, we firstly scanned hard and sharp surfaces of randomly oriented $TiO_2$ nanostructures. After such a procedure, tips' radius was typically in the range of 20-45 nm and can be used under 2500 nN load reproducibly. This problem does not appear for diamond tips. For optical measurements, squares with lateral sizes from 10 to 150 µm were densified with the resolution tuned so that the interpixel distance was less than the tip radius.

AFM lithography was carried out in two ways:

The first one employed capabilities of NT-MDT NEXT II AFM equipped with HybriD mode which physically is equivalent to PeakForce tapping, operating in another frequency range. Controller and software allowed lithographic patterning in various regimes, including the grayscale patterning of raster images uploaded in the software.

The second approach was performed with a SmartSPM AFM manufactured by AIST-NT (currently produced by HORIBA Scientific). We used silicon cantilevers with single crystal diamond tips (D300, SCDprobes).[60] These probes provide stable long-term indentation of even hard materials.[61] In this case, we used standard tapping mode. To ensure the absence of lateral deformation during indentation, we set a high initial amplitude of A = 150 nm and a regime in which a high level of Z-feedback gain caused the self-oscillations of the Z-scanner. These z-scanner oscillations led to a 100% modulation of the probe's oscillations with a frequency of about 300 Hz. The indentation depth was regulated by set-point values (50-80 % of initial amplitude) and each line's number of passes (1-5). For delicate AFM imaging of SWCNT films we used a vertical mode[62] with A about 10 nm, and the probes with grown spikes.[63]

*Samples characterization*: For UV-vis-NIR transmission spectroscopy (**Figure 3A**) and THz spectroscopy (**Figure S2**) $T_{600}$ ~ 45 % SWCNT film was firstly transferred onto Si/SiO$_2$ (300 nm, 6x6 mm$^2$) wafer. The surface was patterned by an AFM with a 20x20 square grid of 150x150 µm$^2$ each square size within 14 days. Afterward, the polyethylene hoop on an aluminum ring (D = 12 mm) was manually pressed to the substrate surface, and the whole film was detached from the silicon dioxide surface onto polyethylene. Samples used for Raman spectroscopy (**Figure 3C-F**) were represented by 50x50 µm$^2$ squares densified in $T_{600}$ ~ 45 % SWCNT film transferred onto the surface of Si/SiO$_2$ (300 nm thick) substrate. Samples used for reflection coefficient measurements shown in **Figure 4A-C** and **E** were SWCNT films with $T_{600}$ ~ 45 % (approximately 300 nm thick) on the surface of fused silica. 50x50 µm$^2$ squares were patterned with the help of AFM under different force loads. For thickness series (**Figure 4D**), SWCNT films with different deposition times (from 1 to 10 min with 1-min interval) and different $T_{600}$ respectively were synthesized and transferred on the surface of fused silica. For AFM lithography, $T_{600}$ ~ 45 % films were transferred onto the surface of fused silica and Si/SiO$_2$. For MIR measurements (Supplementary Materials) $T_{600}$ ~ 45% film was transferred onto the surface of CaF$_2$ and patterned by 100x100 µm$^2$ squares at four different loads: 500, 1000, 2000 and 4000 nN.

*Helium Ion Microscope (HeIM) imaging*: Secondary electron images of SWCNT films were recorded using the Zeiss Orion Plus helium ion microscope at an acceleration voltage of 30 kV. The ion-induced electron emission allows imaging morphological details, similarly to a secondary electron microscope. Compared to electron microscopy, the reduced penetration and scattering of helium ions allow a better spatial resolution down to ~ 0.5 nm. The chamber pressure was 3x10$^{-7}$ Torr during imaging, and the typical ion beam current was 1.3 pA.

*Helium Ion Microscope (HeIM) imaging*: Secondary electron images of SWCNT films were recorded using the Zeiss Orion Plus helium ion microscope at an acceleration voltage of 30 kV. The ion-induced electron emission allows imaging morphological details, similarly to a secondary electron microscope. Compared to electron microscopy, the reduced penetration and scattering of helium ions allow a better spatial resolution down to ~ 0.5 nm. The chamber pressure was $3\times10^{-7}$ Torr during imaging, and the typical ion beam current was 1.3 pA.

*Scanning Electron Microscopy*: We used Jeol JSM-7001F at acceleration voltages from 10 to 30 kV.

*Optical Spectroscopy*: UV-vis-NIR spectrophotometer Perkin Elmer Lambda 1050 was utilized for optical range transmittance spectra capture. The Bruker Vertex 80V equipped with a microscope was used for obtaining NIR-MIR-FIR transmissivity spectra. Room temperature measurements of THz spectra and FIR spectra were performed on a Teraview TPS Spectra 3000 time-domain THz spectrometer and Bruker Vertex V80 FTIR spectrometer, respectively. UV reflection was measured with a custom micro-spectrometer using frequency-doubled Ti:Sapphire laser as a light source. UV light was loosely focused onto the sample using an underfilled 20x plan-fluor objective lens. The illumination spot diameter at the sample was approximately 15 µm. Spectra of the reflected radiation were captured with the Solar S100 spectrometer providing 1 nm spectral resolution. Scattering spectra were measured in a transmitted-light microscopy setup with Köhler illumination. Light from a halogen lamp was focused to the back-focal plane of a plan achromat objective with NA of 0.25. Transmitted and scattered light beams were collected with a water-immersion plan apochromat objective with numerical aperture of 1.2. A pair of lenses after the objective were used to create an image of the back-focal plane of the collecting objective. A central stop placed at the position of the back-focal plane image blocked the directly transmitted light and allowed only the scattered radiation to pass. The latter was then coupled to a multimode fiber with an achromatic lens and directed to an Ocean Optics USB4000 spectrometer. A detailed description of the setup can be found elsewhere.[64]

*Raman Spectroscopy*: Raman spectroscopy was performed with a confocal scanning Raman microscope Horiba LabRAM HR Evolution (HORIBA Scientific, France). All the measurements were carried out using: linearly polarized excitation at a wavelength of 532 nm, a diffraction grating with 1800 grooves per millimeter, and an objective with the magnification of 100x and N.A. = 0.90 (Olympus Microscope Objective M Plan N, Japan). The laser spot size was ~ 0.5 µm. To register a significant signal-to-noise ratio, the unpolarized detection was used. Not to initiate any damage to the samples, the Raman spectra were recorded with the laser power set to 100 mW but passing through the neutral density filter set to 5%: no sample damage was observed during the measurements. For statistics, the spectra of each sample were collected at least in 3 different points of the sample: the standard deviation between the spectra was negligible. The acquisition time of 1 sec at each point was utilized with the accumulation of at least 5 times per point.

*Mesoscopic Distinct Element Method and Simulation Details*: In MDEM representation, a (10,10) SWCNT is divided into distinct elements where each of them lumps a finite number of atoms. The distinct elements in direct contact interact by means of the enhanced vector model potential, which describes the resistance to stretching, shear, bending, and torsional deformation displacements of the SWCNT portion of length T and radius r as an Euler–Bernoulli beam. Distinct elements located on different SWCNTs interact through a coarse-grained anisotropic vdW potential optimized to provide realistic, smooth sliding of aligned SWCNTs. This athermal model is augmented with an energy dissipation model optimized to capture the room temperature sliding and aggregation (zipping) of SWCNTs.

For the film simulations, the starting structure was a computer-generated network containing 1733 straight (10,10) SWCNTs each with 332 nm length, "grown" from "seeds" randomly placed in a $333 \times 333 \times 140$ nm$^3$ cuboid. The total number of distinct elements used to represent the film is 424585. The SWCNT orientations were distributed isotopically in the SWCNT xy plane and uniformly distributed out-of-plane within the angle made by vectors with the xy plane within the borders of 0° and 11°. Then the film was subjected to relaxation for 100 ns with a 20 fs time step. While the film's simulated portion's lateral directions are subjected to periodic boundary conditions, the upper and lower sides form free surfaces. The stable, relaxed film has a height $h_0$ of 127 nm and a density of 0.13 g/cm$^3$, similar to the values of the pristine films collected directly from the floating catalyst (aerosol) chemical deposition reactor.

To further simulate the indentation process, the upper film surface was interacting through a repulsive potential[30] with an infinitely rigid nanoindenter, moving at the speed of 10 m/s. The entire lower surface was placed in interaction with one stationary planar plate. The stress (s) is defined as the sum of forces exerted by the indenter onto the interacting distinct elements divided by the 20x20 nm$^2$ cross-sectional area of the indenter.

The MDEM simulations shown in **Figure S1** predict that the local densification mechanism demonstrated in the main text on pristine low-density films, would also be applicable to pre-densified films. The maximum stress level is similar to the one applied to the pristine film case.

**Supporting Information**

Supporting Information is available from the Wiley Online Library or from the author.

**Acknowledgements**

Authors are grateful to Ms. Anastasiia Grebenko for assistance with sample preparation and Mr. Andrey Starkov for graphical illustrations and engineering assistance. Authors also thank Mr. Anton Bubis for the help with SEM images, prof. Sergey Zaitsev and Dr. Alexandr Svintsov for calculating the CGH. A.K.G. and A.G.N acknowledge support from RFBR grant 19-32-90143. E.M.Kh. and D.V.K. acknowledge Russian Science Foundation grant No. 20-73-10256 (synthesis of SWCNTs and optical measurements). Y.G.G. acknowledge support from RFBR grant 18-29-20032. The Authors acknowledge the use of computational resources of the Skoltech CDISE supercomputer Zhores.[65] I.O. acknowledges support from RFBR grant 18-29-19198. A.P.Ts. acknowledges the EDUFI Fellowship (No. TM-19-11079) from the Finnish National Agency for Education and the Magnus Ehrnrooth Foundation (the Finnish Society of Sciences and Letters) for personal financial support. This work is supported by the Council on grants of the President of the Russian Federation grant number НШ-1330.2022.1.3.

**Author Contributions**

A.K.G. performed experimental design and data processing, performed all densification experiments and sample preparation for optical measurements. G.D. performed MDEM simulations and analysis, S.S.Z. made THz-FIR measurements, E.M.K. and D.V.K. synthesized SWCNT films. A.P.Ts. did Raman mapping, B.A. did UV measurements and scattering investigation, Y.G. analyzed optical properties of diffraction gratings and CGH A.G.T. performed AFML in hard tapping mode, T.D., A.G.N. and I.O. supervised the research. M.L., H.H. and L.C.F. did HeIM measurements, A.K.G., G.D, T.D, A.G.N and V.P. wrote the manuscript. All authors commented on the manuscript.

Atomic Force Microscopy lithography is used to create an optically active medium in the SWCNT network by precise control of its density

*Artem K. Grebenko\*, Grigorii Drozdov, Yuriy G. Gladush, Igor Ostanin, Sergey S. Zhukov, Eldar M. Khabushev, Alexey P. Tsapenko, Dmitry V. Krasnikov, Boris Afinogenov, Alexei G. Temiryazev, Viacheslav V. Dremov, Traian Dumitricã\*\*, Mengjun Li, Hussein Hijazi, Vitaly Podzorov, Leonard C. Feldman, Albert G. Nasibulin\*\*\**

**Local ultra-densification of single-walled carbon nanotube films: modeling and experiment**

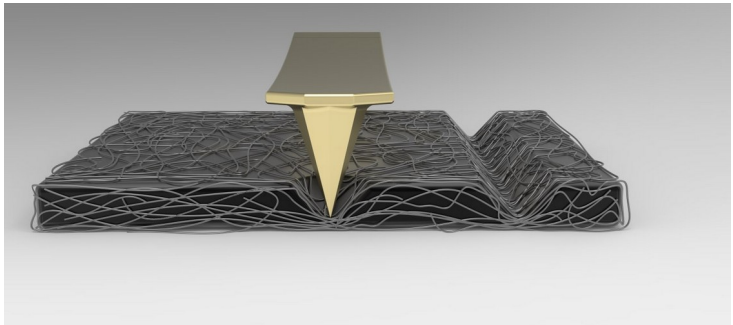

Supporting Information

**Local ultra-densification of single-walled carbon nanotube films: modeling and experiment**

*Artem K. Grebenko\*, Grigorii Drozdov, Yuriy G. Gladush, Igor Ostanin, Sergey S. Zhukov, Eldar M. Khabushev, Alexey P. Tsapenko, Dmitry V. Krasnikov, Boris Afinogenov, Alexei G. Temiryazev, Viacheslav V. Dremov, Traian Dumitricã\*\*, Mengjun Li, Hussein Hijazi, Vitaly Podzorov, Leonard C. Feldman, Albert G. Nasibulin\*\*\**

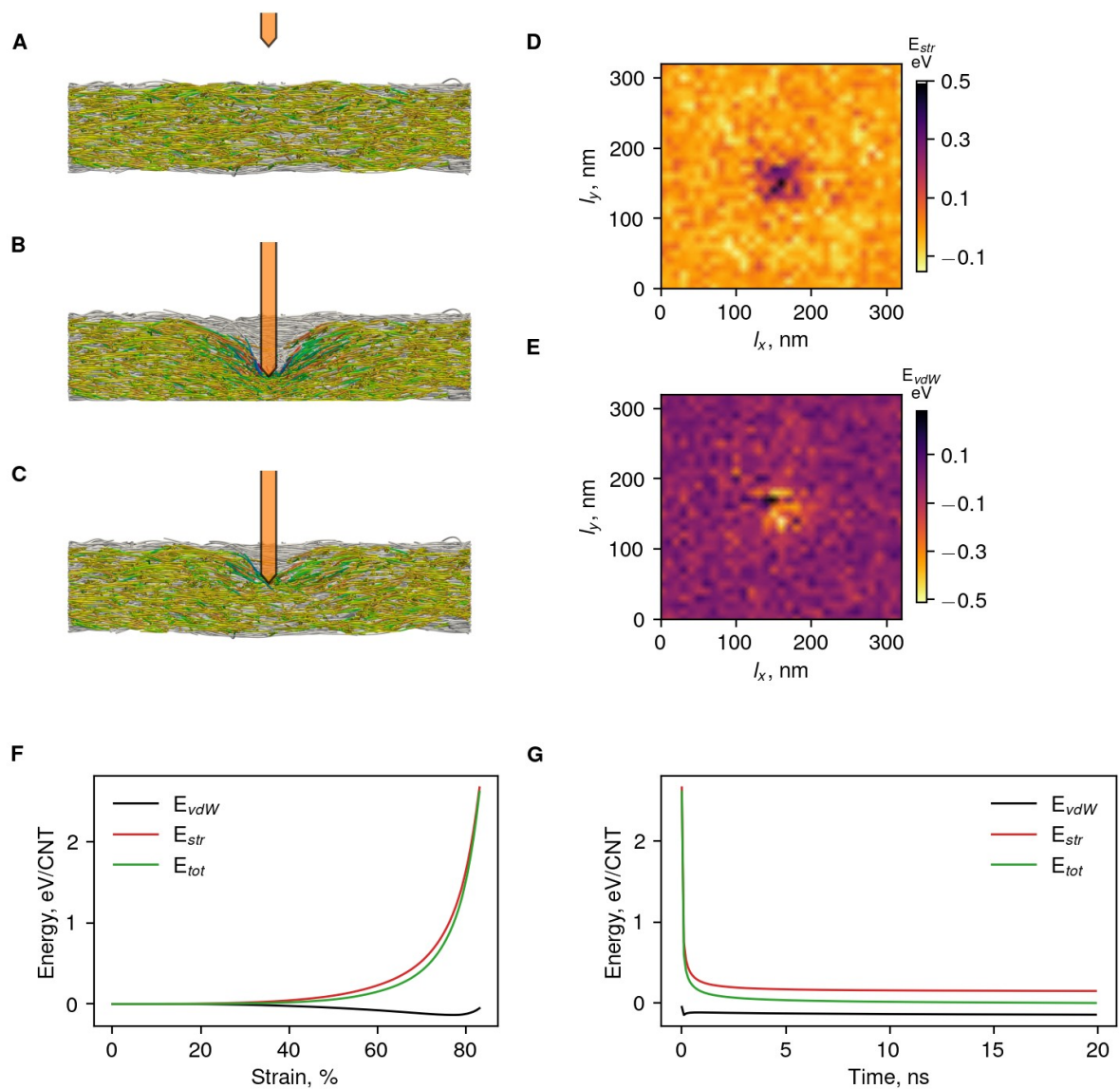

**Figure S1. MDEM simulations of indenter-based densification of a pre-densified film.** (**A**) Initial film of 0.22 g/cm³ density and 75 nm thickness. (**B**) Indentor maximum penetration and (**C**) Retraction stages**.** Maps of (**D**) strain and (**E**) vdW energy changes for the locally-densified film with respect to the initial one. Evolution of film energy during (**F**) push and (**G**) recovery (complete pull out).

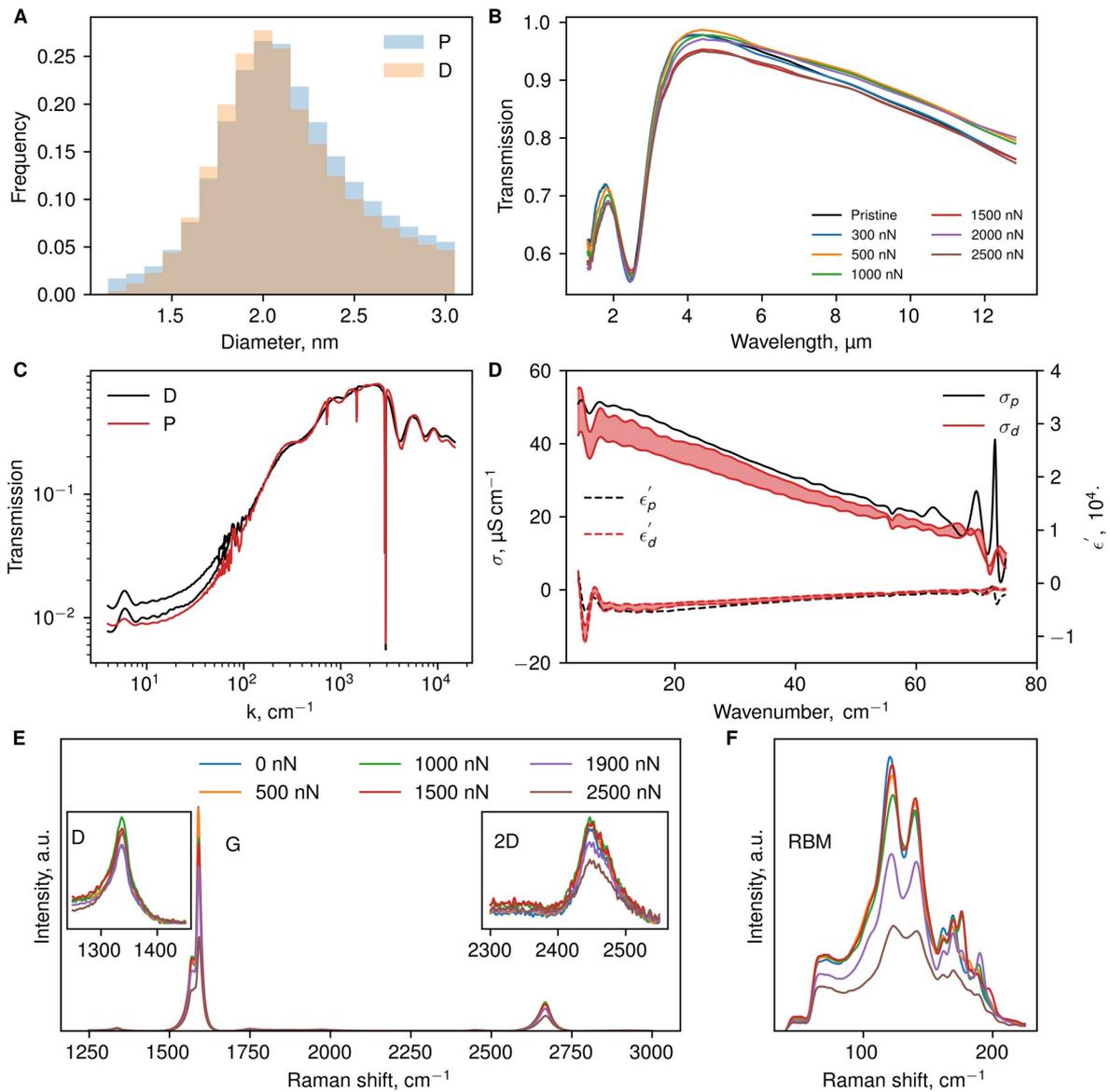

**Figure S2. Supplementary information on THz and Raman spectroscopy.** (**A**) Nanotube diameter distribution calculated from UV-vis transmission spectra indicated in **Figure 3**. (**B**) MIR transmission spectrum of SWCNT film (CaF$_2$ substrate) regions densified with different load force values indicated in the legend. (**C**) THz-FIR spectrum of pristine (P) and densified (D, 2500 nN) 3x3 mm sample on polyethylene substrate. (**D**) THz spectroscopy data indicating σ and ε' for the same set of samples as for panel **C**. The faint red area shows the measurement error caused by the sample's orientation relative to the k vector of the incident wave. (**E**) Raman spectra of densified regions in a SWCNT film on Si/SiO$_2$ substrate. (**F**) RBM region for the same nanotubes as in panel **E**. Legend is the same for panels **E** and **F**.

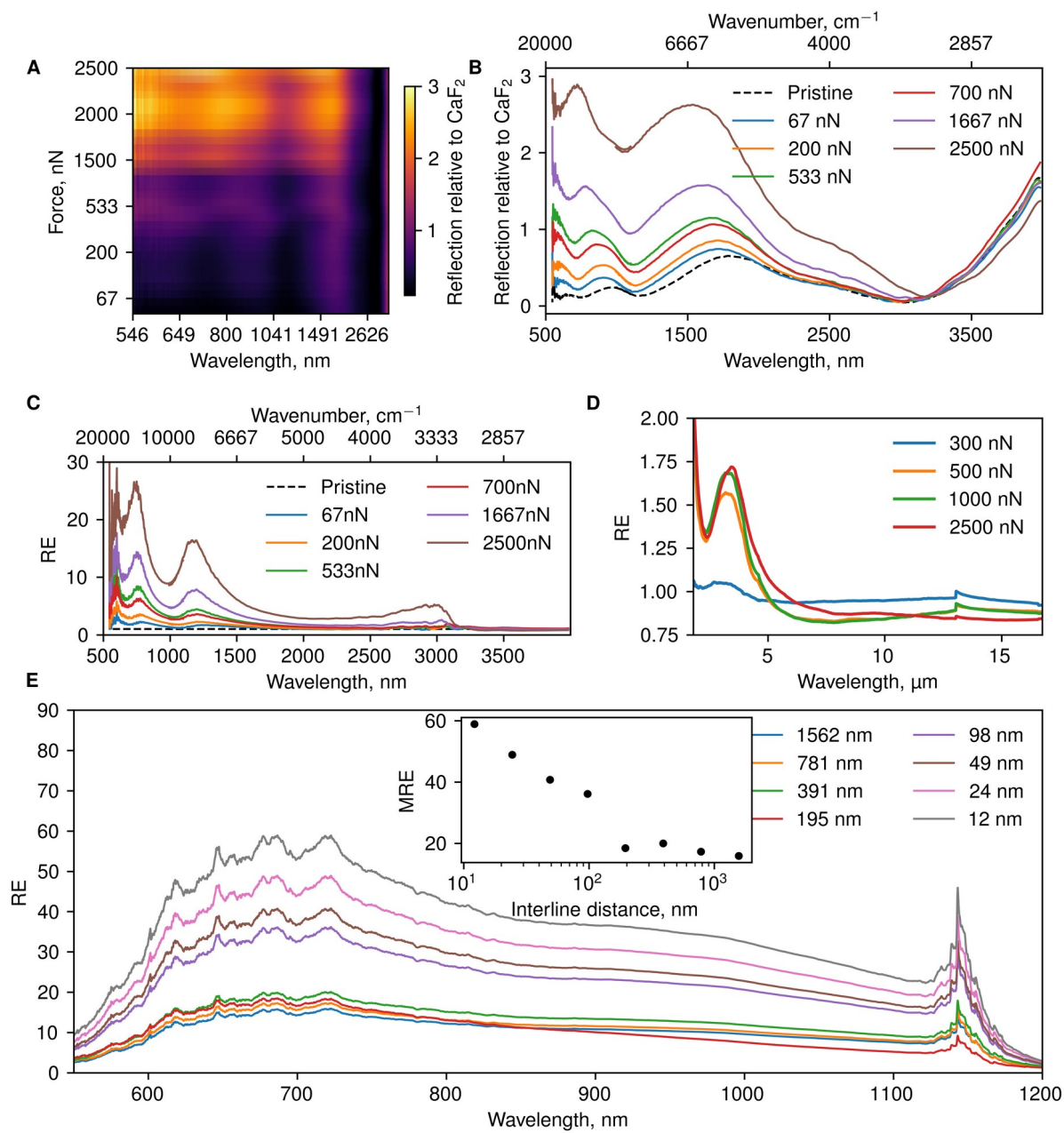

**Figure S3. Supplementary information on the reflection enhancement spectroscopy.** (**A**) Reflection coefficient normalized per reflection of the quartz substrate in the form of a 2D map. (**B**) The same data presented in the form of individual spectra. (**C**) Map from **Figure 4A** showed in the form of individual spectra (RE - reflection enhancement relative to the pristine SWCNTs). (**D**) Reflection enhancement spectrum for MIR region captured from the densified regions in SWCNT film on $CaF_2$ substrate. (**E**) Reflection enhancement spectra for SWCNT film densified with different interline distances during densification procedure.

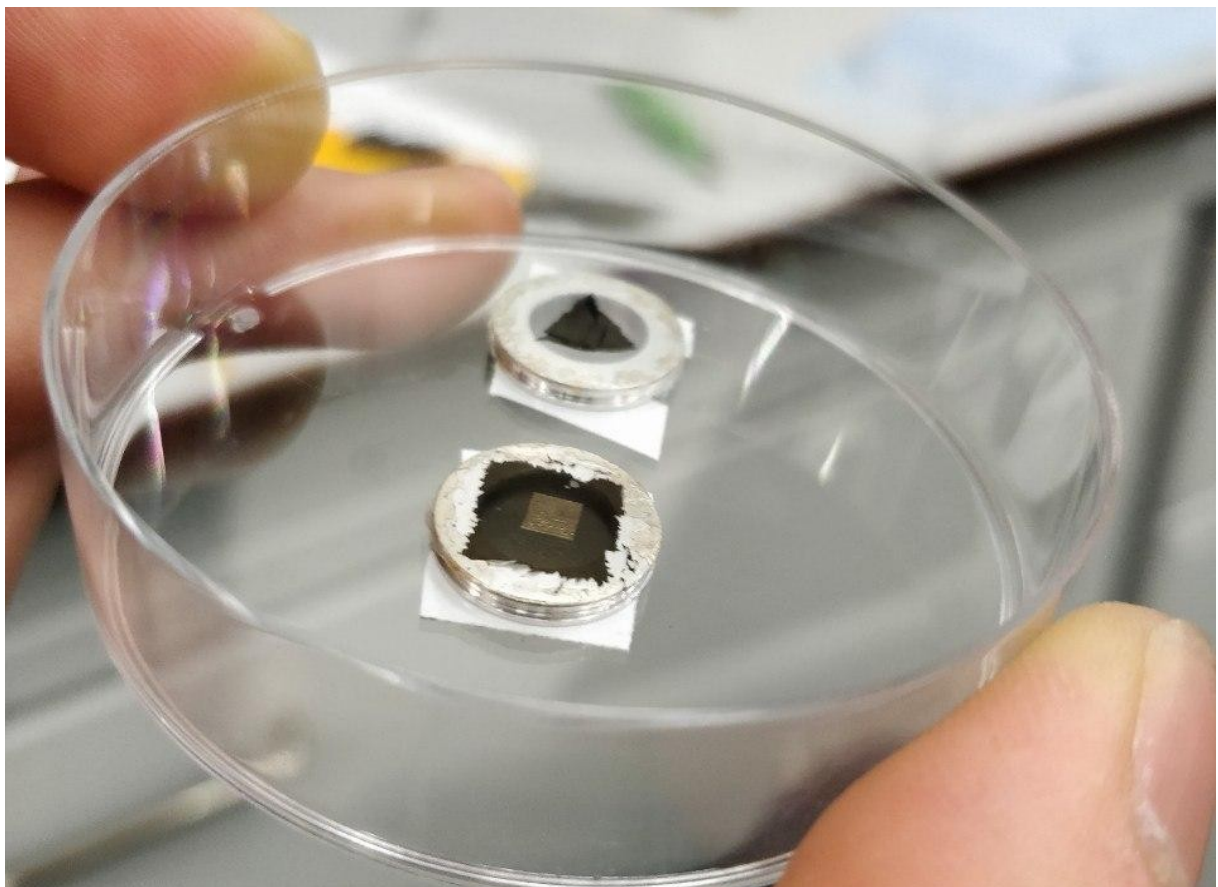

**Figure S4. Large area densified SWCNT film.** Photograph of 3x3 mm2 densified region in a SWCNT film transferred to the polyethylene substrate from silicon.